\documentstyle[preprint,aps]{revtex}
\begin{document}
\title{A Hamiltonian for the description \\
of a non-relativistic spin-1/2 free particle}
\author{S. Bruce}
\address{Department of Physics, University of Concepcion, P.O. Box 160-C, Concepcion,%
\\
Chile}
\date{\today}
\maketitle

\begin{abstract}
We propose a Hamiltonian for a nonrelativistic spin 1/2 {\it free} particle
(e.g. an electron) and find that it contains information of its internal
degrees of freedom in the rest coordinate system. We comment on the
dynamical symmetry associated with the electron {\it Zitterbewegung}.

\end{abstract}

\newpage

The derivation of the Dirac equation\cite{DI} begins with the attempt to
linearize the Klein-Gordon equation for the free particle, which is
quadratic in the energy-momentum operator $\widehat{p}^{\mu }=i\hbar
\partial /\partial q_{\mu }$ (we use the metric signature $g(+---)),$ i.e., $%
\left( \left( c\widehat{p}^{0}\right) ^{2}-c^{2}\widehat{{\bf p}}%
^{2}-m_{0}^{2}c^{4}\right) \Phi (q)=0,$ where $c$ is the velocity of light, $%
m_{0}$ is the mass of the particle, $q=(q_{,}^{0}q^{i}{\bf )}=(ct,{\bf q)},$
and $\Phi $ a Lorentz scalar wave function of $q$. It has a $\left( c%
\widehat{p}^{0}\right) ^{2}=-\left( \hbar c\right) ^{2}\partial
^{2}/\partial q_{0}^{2}$ term: this leads to a continuity equation with a
probability density containing $\partial /\partial q_{0}$, and hence to
negative probability. To solve these problems we require an equation linear
in $\widehat{p}=\left( \widehat{p}^{0},\widehat{p}^{i}\right) =\left( E/c,%
\widehat{{\bf p}}\right) {\bf .}$ In fact, a Lorentz invariant wave equation
can be constructed\cite{DI} in the form $\left( \widehat{p}^{0}-{\bf \alpha
\cdot }\widehat{{\bf p}}{\bf -}m_{0}c\beta \right) \Psi (q)=0.$ The operator
in this equation contains $q_{\mu }$ , $\mu =0,1,2,3,$ only in derivative $%
\partial /\partial q_{\mu }.$ Notice that this implies the existence of a
reference system with a well defined {\it origin}, where the wave equation
has been written down. On the other hand, since ``all space-time points are
equivalent'', ${\bf \alpha }$ and $\beta $ ``must not involve $q$'' \cite{DI}%
.

It is well known\cite{LL} that in nonrelativistic quantum mechanics it is
possible to construct a Dirac-like {\it wave equation} to describe a spin
1/2 particle: the L\'{e}vy-Leblond equation (LE). However, from the LE it is
not possible to define a Hamiltonian for the description of a given quantum
system since the time derivative in the wave function is multiplied by a 
{\it singular} $4\times 4$ matrix. On the other hand we know that the Pauli
equation does not describe {\it Zitterbewegung}.

In this article we want to partially remedy these problems by proposing a
Hamiltonian for a spin 1/2 particle (e.g. an electron) which both it is
intrinsically nonrelativistic and contains information of its spin in the
rest system. The ``price'' we have to pay is that we need to incorporate a
quadratic term in $\widehat{{\bf p}}$ in the Hamiltonian. As this is
preliminary work, we shall concern ourselves only with the case of the free
particle.

To begin with, we consider the classical expression for the energy of a
nonrelativistic free particle, including its rest energy: 
\begin{equation}
E\left( {\bf p}\right) =m_{0}c^{2}+\frac{{\bf p}^{2}}{2m_{0}}.
\end{equation}
If we square this equation we obtain 
\begin{equation}
E^{2}\left( {\bf p}\right) =m_{0}^{2}c^{4}+c^{2}{\bf p}^{2}+\frac{\left( 
{\bf p}^{2}\right) ^{2}}{4m_{0}^{2}}.  \label{2}
\end{equation}
We observe that the first two terms of Eq.(\ref{2}) (taking them together)
resemble the expression for the squared of the free particle Dirac equation
(the Klein-Gordon equation). In quantum mechanics, the ``square root'' of (%
\ref{2}), for a {\it strictly} nonrelativistic spin 1/2 particle, can be
written now as the Hamiltonian operator 
\begin{equation}
\widehat{H}=c{\bf \alpha \cdot }\widehat{{\bf p}}{\bf +}m_{0}c^{2}\beta
-i\beta \gamma _{5}\frac{\left( {\bf \alpha \cdot }\widehat{{\bf p}}\right)
^{2}}{2m_{0}},  \label{3}
\end{equation}
where ${\bf \alpha }_{i}{\bf ,}$ $\beta $ are the Dirac matrices, and $%
\gamma _{5}\equiv \gamma _{0}\gamma _{1}\gamma _{2}\gamma _{3}$ in the Dirac
representation\cite{GR}. Thus in the process of linearizing Eq.(\ref{2}), we
pick up a quadratic term in the momentum $\widehat{{\bf p}}{\bf \ }$in Eq.(%
\ref{3}).

The corresponding Schr\"{o}dinger equation is written as usually: 
\begin{equation}
\widehat{H}\Psi \left( {\bf q},t\right) =i\hbar \frac{\partial \Psi \left( 
{\bf q},t\right) }{\partial t},  \label{4}
\end{equation}
with $\Psi $ the four-spinor.

For a free electron, the velocity operator is given by the Heisenberg
equation 
\begin{equation}
\widehat{v}_{i}=\frac{d\widehat{q}_{i}}{dt}=\frac{i}{\hbar }\left[ \widehat{H%
}{\bf ,}\widehat{q}_{i}\right] =\frac{\widehat{\Gamma }}{m_{0}}\widehat{p}%
_{i}+c\alpha _{i},  \label{6}
\end{equation}
where $\widehat{\Gamma }=-i\beta \gamma _{5}.$ Expression (\ref{6}) states
that the electron possess two types of coordinates: an ``external'' one
proportional to the usual velocity $\widehat{p}_{i}/m_{0}$ operator and an
``internal'' or ``microscopic''\cite{BA} one given by $c\alpha _{i}.$ In
fact we can determine the position operator associated for the electron. To
this end, we observe that 
\begin{equation}
\left\{ \widehat{H},\widehat{v}_{i}\right\} =2\widehat{H}\widehat{v}_{i}+%
\left[ \widehat{v}_{i},\widehat{H}\right] =\frac{2E}{m_{0}}\widehat{p}_{i},
\end{equation}
where $E=+\sqrt{c^{2}{\bf p}^{2}+m_{0}^{2}c^{4}}.$ Thus 
\begin{equation}
\frac{d\widehat{v}_{i}}{dt}=\frac{i}{\hbar }\left[ \widehat{H}{\bf ,}%
\widehat{v}_{i}\right] =\frac{2i}{\hbar }\widehat{H}\left( \widehat{v}_{i}-%
\frac{\widehat{H}^{-1}E}{m_{0}}\widehat{p}_{i}\right) .  \label{8}
\end{equation}
Let us define the operator 
\begin{equation}
\widehat{\eta }_{i}\equiv \widehat{v}_{i}-\widehat{H}^{-1}E\frac{\widehat{p}%
_{i}}{m_{0}}.
\end{equation}
Eq. (\ref{8}) can be regarded as a differential equation $\widehat{v}_{i}$.
Keeping in mind that $\widehat{p}_{i}$ and $\widehat{H}$ are constants of
motion, we see that the $\widehat{\eta }_{i}$ satisfy the differential
equations 
\begin{equation}
\frac{d\widehat{\eta }_{i}}{dt}=\frac{2i}{\hbar }\widehat{H}\widehat{\eta }%
_{i}.
\end{equation}
Solving for $\widehat{\eta }_{i}$ we get 
\begin{equation}
\widehat{\eta }_{i}\left( t\right) =\exp \left( \frac{2i}{\hbar }\widehat{H}%
t\right) \widehat{\eta }_{i}\left( 0\right) .  \label{9}
\end{equation}
As for the coordinate operator, the relation (\ref{9}) can be integrated to
yield 
\[
q_{i}\left( t\right) =q_{i}\left( 0\right) +E\widehat{H}^{-1}\frac{\widehat{p%
}_{i}}{m_{0}}t-\frac{ic\hbar }{2}\widehat{H}^{-1}\exp \left( \frac{2i}{\hbar 
}\widehat{H}t\right) \widehat{\eta }_{i}\left( t\right) . 
\]
For the electron at rest $\left( p_{i}=0\right) ,$ we have that the
operators 
\begin{eqnarray*}
\widehat{H}_{0} &=&m_{0}c^{2}\beta ,\qquad \widehat{S}_{k}=\frac{\hbar }{2}%
\Sigma _{k}\equiv -\frac{\hbar }{2}\alpha _{i}\alpha _{j}, \\
\widehat{P}_{i} &\equiv &m_{0}\left. \widehat{v}_{i}\right| _{{\bf %
p\rightarrow 0}}=m_{0}c\alpha _{i},\qquad \widehat{Q}_{i}\equiv \left. 
\widehat{q}_{i}\right| _{{\bf p\rightarrow 0}}=-\frac{i\hbar }{2m_{0}c}\beta
\alpha _{i},
\end{eqnarray*}
together with $i\gamma _{5},$ $\beta \gamma _{5}$ and $i\beta S_{i},$ form a 
$so(4,2)$ Lie algebra, corresponding to the dynamical symmetry of this
system. Notice that this fact is naturally deduced by using the Dirac
equation, i.e., from a relativistic point of view \cite{BA,BR}.

This work was supported by Direcci\'{o}n de Investigaci\'{o}n, Universidad
de Concepci\'{o}n, through grant \#96.011.019-1.0.

\end{document}